\documentclass[twocolumn,tighten,letter]{aastex631}

\shorttitle{First Space-VLBI of Sagittarius~A$^\ast$}
\shortauthors{Johnson et al.}

\usepackage{newtxmath,newtxtext,xspace,amsmath}

\def\m87{{M87$^\ast$\xspace}}
\newcommand{\sgra}{Sgr~A$^\ast$\xspace}
\newcommand{\ra}{\emph{RadioAstron}\xspace}

\renewcommand{\O}[1]{\mathcal{O}\pa{#1}}

\def\lsim{\mathrel{\raise.3ex\hbox{$<$\kern-.75em\lower1ex\hbox{$\sim$}}}}
\def\gsim{\mathrel{\raise.3ex\hbox{$>$\kern-.75em\lower1ex\hbox{$\sim$}}}}

\defcitealias{PaperI}{Paper~I}
\defcitealias{PaperII}{Paper~II}
\defcitealias{PaperIII}{Paper~III}
\defcitealias{PaperIV}{Paper~IV}
\defcitealias{PaperV}{Paper~V}
\defcitealias{PaperVI}{Paper~VI}
\defcitealias{PaperVII}{Paper~VII}

\defcitealias{Johnson_2018}{\tt J18}

\begin{document}

\title{ First Space-VLBI Observations of Sagittarius~A$^\ast$}

\correspondingauthor{M. D. Johnson}
\email{mjohnson@cfa.harvard.edu}
\author[0000-0002-4120-3029]{Michael~D.~Johnson}
\affiliation{Center for Astrophysics | Harvard \& Smithsonian, 60 Garden St, Cambridge, MA 02138, USA}

\author[0000-0001-9303-3263]{Yuri Y. Kovalev}
\affiliation{Lebedev Physical Institute of the Russian Academy of Sciences, Leninsky prospekt 53, 119991 Moscow, Russia}
\affiliation{Moscow Institute of Physics and Technology,
  Institutsky per. 9, Dolgoprudny, Moscow region, 141700, Russia}
\affiliation{Max-Planck-Institut f\"ur Radioastronomie, Auf dem H\"ugel 69,
53121 Bonn, Germany}

\author{Mikhail M. Lisakov}
\affiliation{Max-Planck-Institut f\"ur Radioastronomie, Auf dem H\"ugel 69, 53121 Bonn, Germany}
\affiliation{Lebedev Physical Institute of the Russian Academy of Sciences, Leninsky prospekt 53, 119991 Moscow, Russia}

\author[0000-0002-1290-1629]{Petr A. Voitsik}
\affiliation{Lebedev Physical Institute of the Russian Academy of Sciences, Leninsky prospekt 53, 119991 Moscow, Russia}

\author{Carl R.~Gwinn}
\affiliation{University of California at Santa Barbara, Santa Barbara, CA 93106-4030, USA}

\author{Gabriele Bruni}
\affiliation{INAF -- Istituto di Astrofisica e Planetologia Spaziali, via Fosso del Cavaliere 100, 00133 Rome, Italy}

\begin{abstract}
We report results from the first Earth-space VLBI observations of the Galactic Center supermassive black hole, \sgra. 
These observations used the space telescope Spektr\nobreakdash-R of the \ra project together with a global network of 20 ground telescopes, observing at a wavelength of $1.35\,{\rm cm}$. Spektr\nobreakdash-R provided baselines up to $3.9$ times the diameter of the Earth, corresponding to an angular resolution of approximately $55\,\mu{\rm as}$ and a spatial resolution of $5.5 R_{\rm Sch}$ at the source, where $R_{\rm Sch} \equiv 2 G M/c^2$ is the Schwarzschild radius of \sgra. 
Our short ground baseline measurements ($\lsim 80\,{\rm M}\lambda$) are consistent with an anisotropic Gaussian image, while our intermediate ground baseline measurements ($100 - 250\,{\rm M}\lambda$) confirm the presence of persistent image substructure in \sgra. 
Both features are consistent with theoretical expectations for strong scattering in the ionized interstellar medium, which produces Gaussian scatter-broadening on short baselines and refractive substructure on long baselines.  
We do not detect interferometric fringes on any of the longer ground baselines or on any ground-space baselines. While space VLBI offers a promising pathway to sharper angular resolution and the measurement of key gravitational signatures in black holes, such as their photon rings, our results demonstrate that space VLBI studies of \sgra\ will require sensitive observations at submillimeter wavelengths.
\end{abstract}

\keywords{ radio continuum: ISM --- scattering --- ISM: structure --- Galaxy: nucleus --- techniques: interferometric --- turbulence}

\section{Introduction}
\label{sec:intro}

Very long baseline interferometry (VLBI) provides the highest direct angular resolution in astronomy, recently culminating in the first image of a black hole \citep{PaperI}. Images of black holes are expected to have a rich interferometric response on long baselines, including small-scale power from both their irregular accretion flows \citep[e.g.,][]{Roelofs_2017,Medeiros_2018,Gelles_2021} and from their distinctive gravitationally lensed features such as the ``photon ring'' \citep[e.g.,][]{Johnson_Ring,Gralla_2020}.  A crucial target for VLBI studies is the Milky Way's nuclear black hole, Sagittarius~A* (\sgra), which subtends the largest angular size of any known black hole. 
The intrinsic and scattered structure of \sgra\ have been intensively studied using VLBI at millimeter and centimeter wavelengths \citep[e.g.,][]{Jauncey_1989,Alberdi_1993,Lo_1998,Krichbaum_1998,Bower_2004,Shen_2005,Doeleman_2008,Lu_2011,Johnson_2018,Issaoun_2019}. A major challenge for these studies is the unusually strong interstellar scattering of \sgra.

\citet{Kellermann_1977} reported early VLBI observations of \sgra\ at $\lambda = 3.8\,{\rm cm}$, which indicated that 25\% of the emission came from a compact component, $\lsim 1\,{\rm mas}$ in size. 
Such a bright, compact component would be detectable using ground-space VLBI with the orbiting telescope Spektr\nobreakdash-R of the \ra program \citep{2013ARep...57..153K}. More recent measurements have not found evidence for this component at centimeter wavelengths \citep[e.g.,][]{Lo_1998,Bower_2004,Lu_2011,Bower_2015,Johnson_2018}. However, \citet{Gwinn_2014} discovered compact image substructure in \sgra\ at $\lambda = 1.3\,{\rm cm}$, with approximately $1\%$ of the total flux density measured on baselines up to $3000\,{\rm km}$ (${\approx}\, 250 \times 10^6 \lambda$, or $0.8\,{\rm mas}$ resolution). This signal is consistent with expected ``refractive substructure'' produced by scattering in the ionized interstellar medium \citep{NG_1989,GN_1989,Johnson_Gwinn_2015}. 
Refractive substructure was also detected in AGN using space-VLBI \citep{2016ApJ...820L..10J,2018MNRAS.474.3523P}.

In this Letter, we examine the interferometric properties of \sgra\ on the much longer baselines, using ground-space VLBI provided by the orbiting telescope Spektr\nobreakdash-R. Launched in 2011, Spektr\nobreakdash-R was located on a highly elliptic orbit, with perigee of $350{,}000\,{\rm km}$ and an orbital period of approximately 9~days. It successfully observed using receivers at four wavelengths, extending from 1.3\,cm to 92\,cm \citep[e.g.,][]{2017MNRAS.465..978P,Kovalev_2020}. Our observations of \sgra\ include projected baselines 4 times the Earth's diameter, providing the sharpest view of this source at centimeter wavelengths and the first using space-VLBI.

\floattable
\begin{deluxetable*}{lccccc}[ht!]
\tablecaption{Observing setup and array for 1.35\,cm observations of \sgra.}
\tablehead{
\colhead{Telescope} & \colhead{Telescope}    & \colhead{Diameter} & \colhead{Bitrate} & \colhead{IFs per}           & \colhead{Observing} \\[-6pt]
\colhead{name}      & \colhead{abbreviation} & \colhead{[m]}      & \colhead{[Mbps]} & \colhead{polarization [MHz]} & \colhead{time [UT]}
}
\startdata
Spektr\nobreakdash-R  &  RA   &  10          & 128  & 2$\times$16 & 13.09.2015 01:30 -- 04:30\\
VLBA St.~Croix     &  SC   &    25          & 2048 & 4$\times$64 & 12.09.2015 23:00 -- 13.09.2015 04:30 \\
VLBA Hancock       &  HN   &    25          & 2048 & 4$\times$64 & 12.09.2015 23:00 -- 13.09.2015 04:30 \\
VLBA North Liberty &  NL   &    25          & 2048 & 4$\times$64 & 12.09.2015 23:00 -- 13.09.2015 04:30 \\
VLBA Fort Davis    &  FD   &    25          & 2048 & 4$\times$64 & 12.09.2015 23:00 -- 13.09.2015 04:30 \\
VLBA Los Alamos    &  LA   &    25          & 2048 & 4$\times$64 & 12.09.2015 23:00 -- 13.09.2015 04:30 \\
VLBA Pie Town      &  PT   &    25          & 2048 & 4$\times$64 & 12.09.2015 23:00 -- 13.09.2015 04:30 \\
VLBA Kitt Peak     &  KP   &    25          & 2048 & 4$\times$64 & 12.09.2015 23:00 -- 13.09.2015 04:30 \\
VLBA Owens Valley  &  OV   &    25          & 2048 & 4$\times$64 & 12.09.2015 23:00 -- 13.09.2015 04:30 \\
VLBA Brewster      &  BR   &    25          & 2048 & 4$\times$64 & 12.09.2015 23:00 -- 13.09.2015 04:30 \\
VLBA Mauna Kea     &  MK   &    25          & 2048 & 4$\times$64 & 12.09.2015 23:00 -- 13.09.2015 04:30 \\
VLA                &  VLA  &   27$\times$25 & 2048 & 4$\times$64 & 12.09.2015 23:00 -- 13.09.2015 04:30 \\
GBT                &  GBT  &   100          & 2048 & 4$\times$64 & 12.09.2015 23:00 -- 13.09.2015 03:00 \\
LBA ATCA           &  ATCA &   6$\times$22  & 1024 & 2$\times$64 & 13.09.2015 02:00 -- 12:00 \\
LBA Mopra          &  MP   &   22           & 1024 & 2$\times$64 & 13.09.2015 02:00 -- 12:00 \\
LBA Ceduna         &  CD   &   30           &  512 & 1$\times$64 & 13.09.2015 02:00 -- 12:00 \\
LBA Hobart         &  HO    &   26           &  512 & 1$\times$64 & 13.09.2015 02:00 -- 12:00 \\
KVN Yonsei         &  KY    &   21           & 1024 & 2$\times$64 & 13.09.2015 07:00 -- 12:00 \\
KVN Ulsan          &  KU    &   21           & 1024 & 2$\times$64 & 13.09.2015 07:00 -- 12:00 \\
KVN Tamna          &  KT    &   21           & 1024 & 2$\times$64 & 13.09.2015 07:00 -- 12:00 \\
Tian Mai           &  T6    &   65           &  512 & 4$\times$32 & 13.09.2015 07:00 -- 12:00 \\
\enddata
\label{tab:array}
\tablecomments{The VLA and ATCA are linked radio interferometers of 27 and 6 antennas, respectively, which were each phased up and coherently summed for these observations. 
All telescopes observed dual circular polarizations except Tian Mai, which observed only left-hand circular polarization (LCP).
All ground telescopes used 2-bit sampling; Spektr\nobreakdash-R used 1-bit sampling. Note that the recorded bandwidth and number of intermediate frequency (IF) bands varied significantly across the array, necessitating a multi-stage correlation and analysis procedure.
}
\end{deluxetable*} 

\begin{figure*}[ht]
  \begin{center}
    \includegraphics[width=0.49\textwidth]{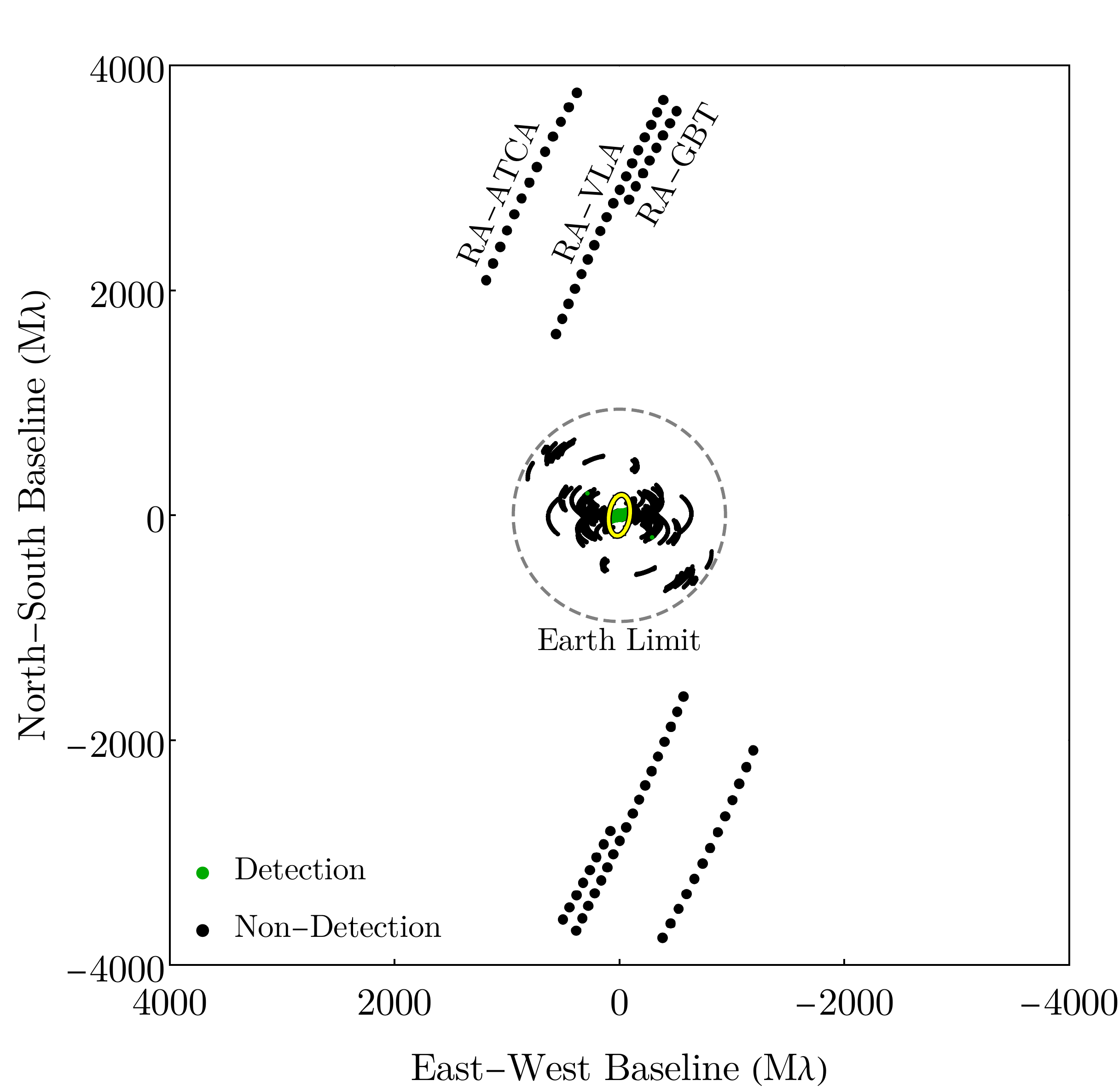}
    \includegraphics[width=0.49\textwidth]{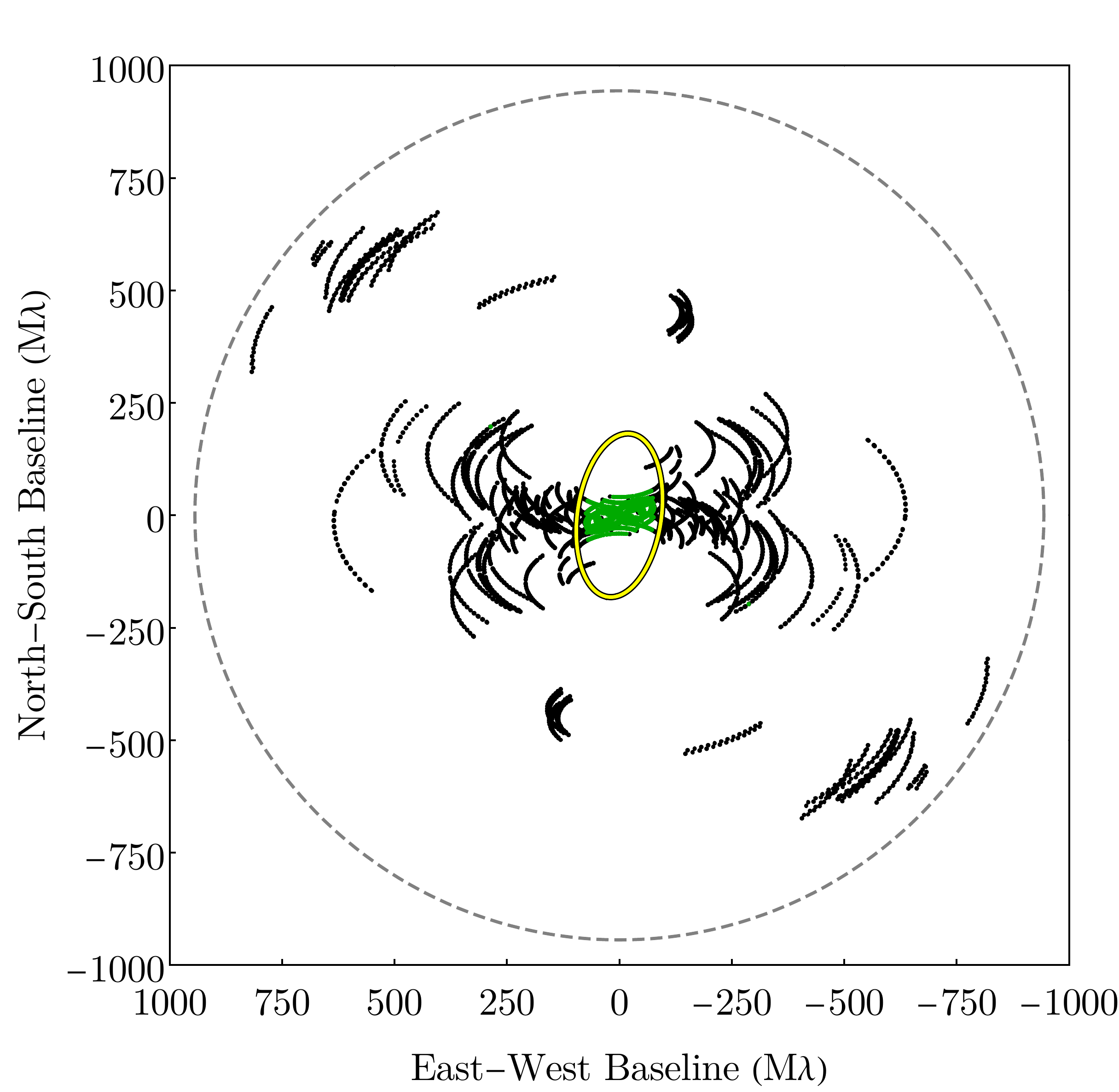}
  \end{center}
\caption{\sgra\ baseline coverage of \ra 1.3~cm observations. The left panel shows the full range of baseline coverage, including only the three most sensitive baselines to Spektr\nobreakdash-R; the right panel shows a zoomed in region with only the ground baselines. Single-baseline detections/non-detections using PIMA are colored green/black. 
We do not detect \sgra\ on any baselines to Spektr\nobreakdash-R. Detections on ground baselines all lie within the region where the scattering kernel is at least $1\%$, denoted by the yellow ellipse.}
\label{fig:uv_coverage}
\end{figure*}

\section{Observations, Correlation, and Processing}
\label{sec:observations}

We observed \sgra\ on September 13, 2015 using Spektr\nobreakdash-R together with a global ground network of 20 telescopes (observing codes RAGS01B and BK193A). To maximize the likelihood of detections on \sgra, which has angular broadening from scattering ${\propto}\,\lambda^2$, our observations used the highest Spektr\nobreakdash-R observing band (a central frequency of 22.2\,GHz). In addition, we selected an observing interval where the Earth-space baselines were predominantly North-South, which is the orientation in which the scattering is both weaker and more poorly constrained.  
\autoref{tab:array} gives details of the array and observing setup. 
The time interval of Spektr\nobreakdash-R observations was determined mainly by its cooling constrains. The Green Bank earth station \citep{2014SPIE.9145E..0BF} was used to collect science data from Spektr\nobreakdash-R.

Correlation was performed using a special version of the software correlator {\tt DiFX} \citep{deller_difx_2011, bruni_difx_2016} that is capable of dealing with Spektr\nobreakdash-R data. 
Due to scheduling limitations, there was no common scan for all stations on the strong fringe calibrator 1730$-$130, however, separate scans for the US and LBA-KVN-T6 parts of the array produced fringes for all ground telescopes. 
Because of the long slew time, Spektr\nobreakdash-R did not observe the calibrator. This decision was supported by an extensive experience with RadioAstron operations; onboard telemetry provided comprehensive information on the technical status of the space telescope.

Due to different recorded bandwidth at different antennas, the correlation was performed in three separate passes to recover as much information as possible:
(i) native 64~MHz channels for all ground stations excluding T6;
(ii) 32~MHz channels for all ground stations;
(iii) $2{\times}16$~MHz channels for baselines between Spektr\nobreakdash-R and GBT, ATCA, and VLA.

Using the VLBI software processing package \texttt{PIMA}\footnote{\url{http://astrogeo.org/pima/}}, we performed baseline-based fringe fitting, which included determination of the phase acceleration term to statistically separate observations with significant interferometric signal and noise. To do this, we applied the approach described in \cite{Petrov_2011} and \cite{Kovalev_2020} to estimate the probability of false detection (PFD) for each interferometric fringe obtained. Fringes with $\text{PFD} > 10^{-4}$ were classified as non-detections; for each non-detection, we computed the associated visibility amplitude upper limit.

The ground array dataset with 64-MHz channels (IFs) was also processed in \texttt{AIPS}\footnote{\url{http://www.aips.nrao.edu}} with global fringe fitting to determine single-IF antenna-based solutions. The amplitude was calibrated using \textit{a~priori} gain and system temperature information provided by the observing facilities.
We were able to perform and use the global antenna-based fringe fitting with multi-IF solutions for the US portion of the array due to availability and successful application of instrumental phase-calibration data. With 4 IFs, this procedure improved the SNR of our solutions by up to a factor of two.

For our subsequent analysis and discussion, we use results from both these software packages, relying on PFDs and upper limits from \texttt{PIMA} for baselines without detected fringes, and using the highest possible signal-to-noise ratio of ground antenna-based fringe fitting solutions in \texttt{AIPS}.

\autoref{fig:uv_coverage} shows the baseline coverage of our observations of \sgra. This figure also indicates which baselines gave statistically significant interferometric single-baseline fringes using PIMA. We did not detect fringes on any baselines to Spektr\nobreakdash-R. 

\section{Analysis and Results}
\label{sec:Results}

Analysis of our observation was complicated by usually short coherence times and poor phasing of the VLA. To ensure accurate results that recover as much information as possible, we performed our analysis in several stages, using strong detections on short baselines to recover weak detections on long baselines. In all cases, we compare our measurements with predictions from a physically motivated model for the scattering of \sgra that was developed by \citet{Psaltis_2018}. We use source and scattering parameters from \citet{Johnson_2018} (hereafter, \citetalias{Johnson_2018}), who estimated them by analyzing archival VLBI measurements of \sgra\ extending from 1.3\,mm to 29\,cm wavelength. The predictions of this model have been successfully tested using VLBI including ALMA at 3.5\,mm \citep{Issaoun_2019,Issaoun_2021} and using monitoring observations with the East Asian VLBI Network at 1.35 and 0.7\,cm \citep{Cho_2021}.  Specifically, the \citetalias{Johnson_2018} model has an intrinsic source that is a circular Gaussian with a wavelength-dependent full width at half maximum (FWHM) size $(0.4\,{\rm mas}) \times \lambda_{\rm cm}$, and a scattering screen located 2.7\,kpc from the Earth and 5.4\,kpc from the source (with a corresponding magnification $M \approx 0.53$). The scattering screen is stochastic, with a shallow scattering power-law index of $\alpha = 1.38$,\footnote{The power-law index $\alpha$ corresponds to that of the phase structure function in the inertial range. The 2D power spectrum of phase fluctuations has an index $\beta = \alpha+2$; the angular broadening scales as $\lambda^{1+2/\alpha}$ at short wavelengths and as $\lambda^2$ at long wavelengths. For a review of interstellar scattering, see \citet{Rickett_1990} or \citet{Narayan_1992}.} an inner scale of $r_{\rm in} = 800\,{\rm km}$, and anisotropic angular broadening at a position angle $\phi_{\rm PA} = 81.9^\circ$ with FWHMs along the principal axes $\theta_{\rm maj} = (1.380\,{\rm mas}) \times \lambda_{\rm cm}^2$ and $\theta_{\rm min} = (0.703\,{\rm mas}) \times \lambda_{\rm cm}^2$ (as $\lambda \rightarrow \infty$). 

On baselines that do not significantly resolve \sgra, the visibility function is expected to be well approximated by that of the ensemble-averaged image -- the product of the visibility function of the intrinsic (unscattered) source and the scattering kernel \citep[e.g.,][]{Coles_1987,GN_1989}. For baselines with $b \lsim (1+M) r_{\rm in} \approx 1200\,{\rm km}$, the scattering kernel is well-approximated by an elliptical Gaussian \citep{GN_1989,Psaltis_2018}; for baselines that only partially resolve the intrinsic source, the source visibility function is also well-approximated by an elliptical Gaussian. Thus, we first fit our short-baseline data using an elliptical Gaussian model, described in \autoref{sec::Gaussian_fits}.

Next, we used the Gaussian fit on short baselines to derived amplitude and phase self-calibration solutions that were applied to all baselines. After discarding data that had no self-calibration solution, we coherently averaged the data in two-minute intervals and across all four IFs to maximize the available sensitivity. We analyzed the averaged data on long ground baselines, which were then sufficiently sensitive to detect a $\lsim 1\%$ signal, described in \autoref{sec:long_baseline_analysis}.

Finally, we assessed whether the long-baseline signal was consistent with the expected refractive noise using the \citetalias{Johnson_2018} model. Because refractive substructure is partially quenched by an extended source, this provides a test of both the intrinsic and scattering model parameters. However, the comparison is statistical in nature because the signal from refractive noise is stochastic. Our comparisons are described in \autoref{sec:consistency_check}.

\subsection{Short-Baseline Analysis: Gaussian Fits}
\label{sec::Gaussian_fits}

We first fit a Gaussian model to the short-baseline data. To avoid biasing this fit with non-Gaussian contributions from refractive-noise dominated measurements, we fit only the data on baselines with an expected ensemble-average flux density of at least $2\%$ of the total flux density in the \citetalias{Johnson_2018} model. The longest baseline meeting this criterion had $|\mathbf{u}| = 97.4$\,M$\lambda$ (BR-VLA). We began by coherently averaging the data in frequency (across each IF) and in time (in 2-second intervals). We then applied amplitude and phase self-calibration solutions before coherently averaging the data further, in 2-minute intervals. At this stage, the primary purpose of self calibration is to avoid phase decoherence in the coherent averaging.

Next, we fit an elliptical Gaussian model to the averaged visibilities. We fit the model using diagonalized log closure amplitudes \citep{Blackburn_2020}, with flat model priors on the major axis, minor axis, and position angle. To avoid non-Gaussian errors from low signal-to-noise ratio (S/N) measurements, we only included closure amplitudes with ${\rm S/N} > 5$. Because of the wavelength-dependent image size, we fit data from each IF independently. To perform the fit, we utilized {\tt eht-imaging} \citep{Chael_16} together with the nested sampling package {\tt dynesty} \citep{Speagle_2020}. 
 
\autoref{tab:Gaussian_Fits} shows the results of this Gaussian model fitting. Our data tightly constrain all the model parameters, particularly the major axis and position angle. Note that the fitted uncertainties on these parameters only give the uncertainties from thermal noise; refractive scattering causes an additional ${\sim}1\%$ variation in the major and minor axis sizes, which is comparable to the error budget from thermal noise for the better-constrained major axis. All our fitted values are consistent with the predictions from the \citetalias{Johnson_2018} scattering model.

{
\begin{deluxetable}{l|ccc}[t]
\tablewidth{\textwidth}
\tablecaption{Summary of Gaussian fits to \sgra.}
\tablehead{
\colhead{$\lambda$ [cm]}  & \colhead{$\theta_{\rm maj}$ [$\mu{\rm as}$]} & \colhead{$\theta_{\rm min}$ [$\mu{\rm as}$]} & \colhead{P.A. [deg]}
}
\startdata
1.352 & $2587  \pm 18$   & $1210 \pm 136$     & $81.8 \pm 0.9$ \\
      & $2580$           & $1394$             & $81.9$\\
      \hline 
1.348 & $2572  \pm 17$   & $1313 \pm 115$     & $82.2 \pm 0.8$ \\
      & $2566$           & $1387$             & $81.9$\\
      \hline 
1.344 & $2571  \pm 17$   & $1356 \pm 112$     & $82.2 \pm 0.9$ \\
      & $2552$           & $1381$             & $81.9$\\
      \hline 
1.341 & $2543  \pm 17$   & $1302 \pm 116$     & $82.5 \pm 0.9$ \\
      & $2538$           & $1374$             & $81.9$    
\enddata
\label{tab:Gaussian_Fits}
\tablecomments{Fitted elliptical Gaussian parameters for \sgra\ for each sub-band (for details, see \autoref{sec::Gaussian_fits}). Below each fitted parameter is its corresponding expected value using the \citetalias{Johnson_2018} source and scattering parameters.}
\end{deluxetable} 
}

\begin{figure*}[t]
  \begin{center}
    \includegraphics[width=\textwidth]{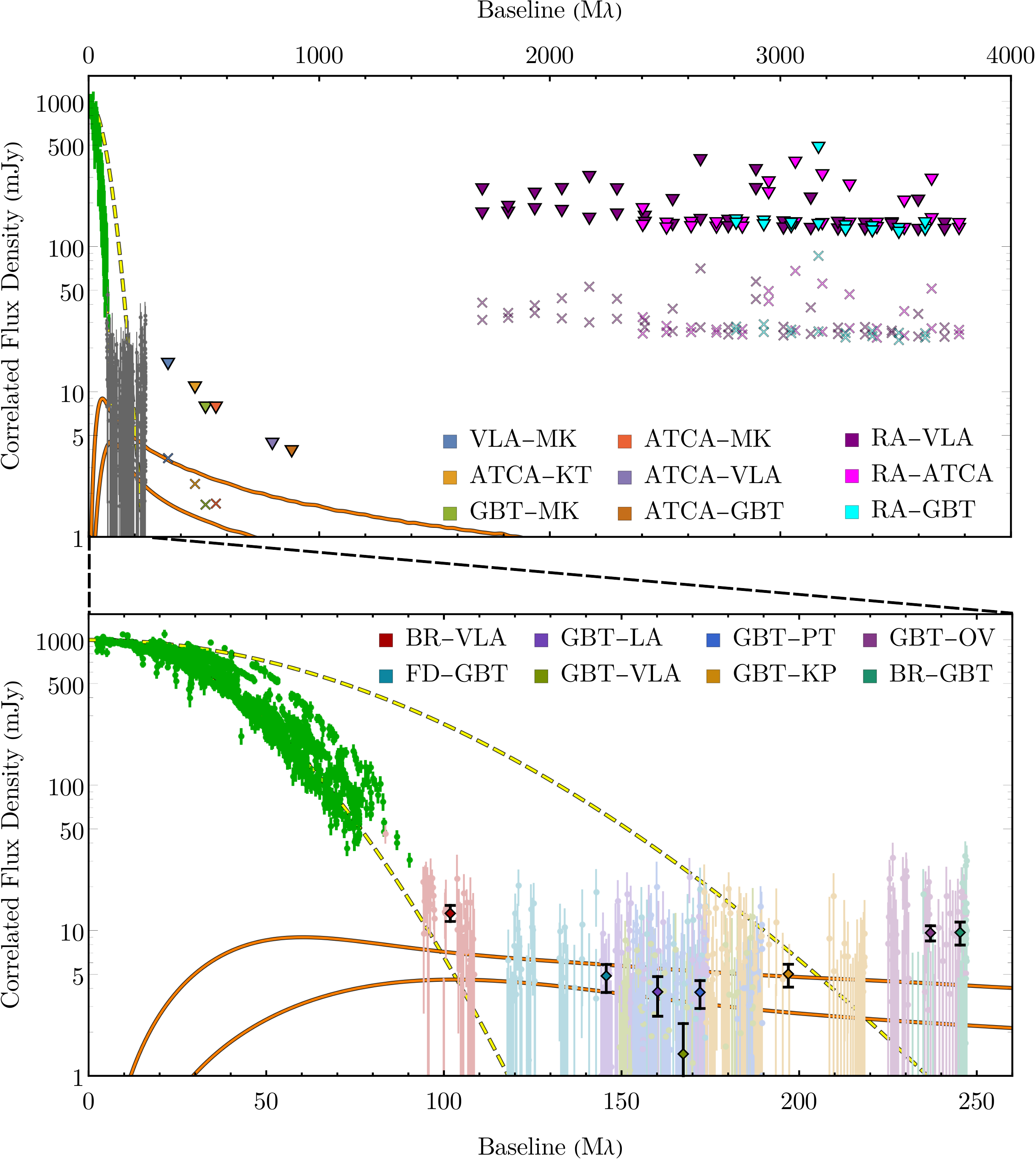}
  \end{center}
\caption{Top: Visibility amplitude as a function of baseline length for \sgra. Green points show strong detections used to derive the Gaussian self-calibration solution. Gray points show the long-baseline visibilities after coherently averaging using the short-baseline self-calibration; for clarity, only visibilities with thermal noise less than $10\,{\rm mJy}$ and $|\mathbf{u}| > 80 \times 10^6$ are shown. Yellow dashed lines shown the expected major (lower) and minor (upper) axes of the ensemble-average image; orange curves show the expected root-mean-square (renormalized) refractive noise along the major (upper) and minor (lower) axes. Colored triangles show the PIMA fringe amplitudes for representative long baselines, including the three most sensitive ground-space baselines; the corresponding thermal noise for each indicated with a cross. None of these fringe amplitudes is a statistically significant detection. Bottom: Zoom-in on ground baselines. Colored points show a subset of the most sensitive long ground baselines; the incoherent average on each baseline is also shown and indicated with a diamond. Uncertainties shown are $\pm1\sigma$.  Both NL-VLA and BR-VLA have significant flux expected in the ensemble-average image, but the remaining long baselines are expected to be dominated by refractive noise.}
\label{fig:radplot_combine}
\end{figure*}

\subsection{Long-Baseline Analysis: Refractive Noise}
\label{sec:long_baseline_analysis}

After performing the Gaussian fits on short baselines, we derived amplitude and phase self-calibration solutions and applied them to all baselines. For the amplitude self-calibration, we used a total flux density of $1\,{\rm Jy}$, which is comparable to the average of historical values at 1.3\,cm \citep[e.g.,][]{Bower_Spectrum}. This choice of normalization does not affect our remaining analysis, which only relies on the fractional visibility amplitudes relative to the total flux density. In addition, we averaged the data across all IFs, to maximize our sensitivity.

\autoref{fig:radplot_combine} shows the resulting visibility measurements, including upper limits on baselines to Spektr\nobreakdash-R (computed by {\tt PIMA}) and highlighting the amplitudes for our most sensitive ground baselines. These baselines are expected to have signals that are dominated by refractive noise; they have $S/N$ up to 8.4 after the final incoherent averaging in time, indicating a reliable detection of image substructure.  
For each figure, we show the expected envelope of the ensemble-average image and the predicted ``renormalized refractive noise'' $\hat{\sigma}_{\rm ref}(\mathbf{u}) = \left\langle \left|\Delta \hat{V}(\mathbf{u}) \right|^2 \right \rangle^{1/2}$, where $\hat{V}$ is the complex visibility function of the source after centering the image and normalizing the total flux density \citepalias[for details, see Appendix~A of][]{Johnson_2018}.

\subsection{Consistency with Expected Properties}
\label{sec:consistency_check}

We now examine whether the long-baselines visibilities are consistent with the expected level of refractive noise for the \citetalias{Johnson_2018} model. For baselines that resolve the ensemble-average image, refractive noise is well approximated as a circular complex Gaussian random variable \citep{Johnson_Narayan_2016}. The visibility amplitude is then drawn from a Rayleigh distribution. The middle 95\% of samples drawn from a Rayleigh distribution fall between $0.18$ and $2.2$ times the mean, $\left \langle \left|\Delta \hat{V}(\mathbf{u}) \right| \right \rangle = \frac{\sqrt{\pi}}{2} \hat{\sigma}_{\rm ref}(\mathbf{u}) \approx 0.89 \hat{\sigma}_{\rm ref}(\mathbf{u})$. All the long-baseline measurements in \autoref{fig:radplot_combine} are, thus, consistent with the expected level of refractive noise. 

We can make more precise comparisons by averaging measurements from different baselines, which sample partially independent elements of the refractive noise. As a simple comparison, we took an incoherent average of all (debiased) visibility amplitudes on baselines longer than $150\times 10^6$ and with thermal noise less than $10\,{\rm mJy}$. This procedure gave an average amplitude of $5.7\,{\rm mJy}$. When then generated a set of 1000 scattered images using the {\tt stochastic-optics} module of {\tt eht-imaging} \citep{Johnson_2016}; we created synthetic observations for each, with baseline coverage and thermal noise matching our observations. For each synthetic observation, we computed the incoherent average on long baselines exactly as was done for the real observation. This procedure gave a median amplitude of $5.4\,{\rm mJy}$, with the inner 95\% of samples falling between $4.3$ and $8.0\,{\rm mJy}$. Hence, the average of our long-baseline measurements is consistent with the predicted range of refractive noise. 

Our non-detections to Spektr\nobreakdash-R are likewise consistent with theoretical expectations for a power-law theory of the scattering and substructure (see \autoref{fig:radplot_combine}). We do not find evidence for any long-baseline visibilities above the expected refractive noise floor, $\sigma_{\rm ref} \sim 1\,{\rm mJy} \times \left( |\mathbf{u}|/{10^9} \right)^{-0.69}$. 
In particular, this refractive noise is ${\sim}2$ orders of magnitude below the most stringent upper limits of ${\sim}150\,{\rm mJy}$ on baselines to Spektr\nobreakdash-R.

An unexpected feature of our long ground baseline measurements is that they are quite close in amplitude to those of \citet{Gwinn_2014} \citepalias[see Figure~6 of][]{Johnson_2018}, which were taken on March 7, 2014. Namely, both observations measure visibility amplitudes of ${\sim}4\,{\rm mJy}$ on baselines from 150-200\,M$\lambda$, with a rise to ${\sim} 8\,{\rm mJy}$ on the ${\sim}250\,{\rm M}\lambda$ baselines GBT-BR and GBT-OV. These observations are separated by 554~days, while the expected correlation timescale of the refractive noise is a few months (for a characteristic scattering velocity of $50\,{\rm km/s}$). Thus, we expect that the similarity is simply a chance alignment; if future measurements show that this signal is persistent, it would either indicate a much slower scattering velocity or that the measurements are sensitive to the ensemble-average properties of the scattering rather than of refractive substructure.

\section{Summary}
\label{sec:Summary}

Space-VLBI is an exciting frontier for black hole astrophysics, with the potential to resolve the gravitationally lensed ``photon rings'' of nearby supermassive black holes \citep{Johnson_Ring}, to measure the masses of \emph{thousands} of supermassive black holes via their ``shadow'' diameters \citep{Pesce_2021}, and to track the orbits of many supermassive black hole binaries \citep{DOrazio_2018}. 
Proposed mission concepts to observe \sgra\ have focused on exploring configurations that allow for rapid baseline sampling to reconstruct movies \citep[e.g.,][]{Palumbo_2019,Johnson_Astro2020} as well as enabling observations at submillimeter wavelengths for which interstellar scattering is sharply reduced \citep[e.g.,][]{Roelofs_2019,Kudriashov_2021}. 

Space VLBI at longer wavelengths provides crucial input for these designs and gives firm system requirements.  In this Letter, we have presented the first space-VLBI of \sgra, observing at $\lambda = 1.35\,{\rm cm}$ using Spektr\nobreakdash-R and 20 ground antennas. Our short ground baselines are well fit by an elliptical Gaussian image, with parameters matching those of historical measurements. Our long ground baselines confirm the presence of persistent small-scale structure in the scattered image of \sgra, originally discovered by \citet{Gwinn_2014}, at a level that is also consistent with predictions for refractive interstellar scattering. We do not detect interferometric fringes on any baselines to Spektr\nobreakdash-R.  

Our work highlights the severe challenges for observing \sgra\ with space-VLBI. Because of diffractive interstellar scattering, improving the angular resolution achievable from the ground will likely require observations  with $\lambda \lsim 2\,{\rm mm}$, while detecting features such as the lensed photon ring will require $\lambda \lsim 0.8\,{\rm mm}$.

\begin{acknowledgments}
We thank Eduardo Ros for useful comments on the manuscript, the anonymous ApJL referee for comments that clarified the manuscript, Amy Mioduszewski for consultations regarding VLA observations in the VLBI mode, and Rebecca Azulay for correlator support at MPIfR. The \ra project was led by the Astro Space Center of the Lebedev Physical Institute of the Russian Academy of Sciences and the Lavochkin Scientific and Production Association under a contract with the State Space Corporation ROSCOSMOS, in collaboration with partner organizations in Russia and other countries.
The National Radio Astronomy Observatory and the Green Bank Observatory
are facilities of the National Science Foundation operated under cooperative agreement by Associated Universities, Inc.
The Australian Long Baseline Array is part of the Australia Telescope National Facility which is funded by the Australian Government for operation as a National Facility managed by CSIRO.
This research is based on observations correlated at the Bonn Correlator, jointly operated by the Max Planck Institute for Radio Astronomy (MPIfR), and the Federal Agency for Cartography and Geodesy (BKG). MDJ thanks the National Science Foundation (AST-1716536, AST-1935980) and the Gordon and Betty Moore Foundation (GBMF-5278) for financial support of this work. YYK is supported in the framework of the State project ``Science'' by the Ministry of Science and Higher Education of the Russian Federation under the contract 075-15-2020-778.
\end{acknowledgments}

\facilities{RadioAstron Space Radio Telescope (Spektr\nobreakdash-R), VLBA, GBT, VLA, LBA, KVN}

\software{
AIPS \citep{Greisen_2003},
eht-imaging \citep{Chael_2016},
PIMA \citep{Petrov_2011}
}

\bibliography{SgrA_RadioAstron.bib}{}
\bibliographystyle{aasjournal}

\end{document}